\documentclass[twocolumn,showpacs,prc,aps,amsmath,amssymb]{revtex4}

\usepackage{graphicx}
\usepackage{dcolumn}
\usepackage{bm}

\begin{document}

\title{Shell Model Analysis of the Neutrinoless Double Beta Decay of $^{48}$Ca}

\author{Mihai Horoi}
 \email{horoi@phy.cmich.edu}
 \homepage{http://www.phy.cmich.edu/people/horoi}
\affiliation{Department of Physics, Central Michigan University, Mount Pleasant, Michigan 48859, USA}

\author{Sabin Stoica}
\affiliation{Horia Hulubei National Institute for Physics and Nuclear Engineering (IFIN-HH) \\
407 Atomistilor, Magurele-Bucharest 077125, Romania}

\date{\today}

\begin{abstract} 
The neutrinoless double beta
($0\nu\beta\beta$) 
decay process could provide crucial information to 
determine the absolute scale of the neutrino masses, and it is the only
one that can establish whether neutrino is a Dirac or a Majorana
particle. A key ingredient for extracting the absolute neutrino
masses from $0\nu\beta\beta$ decay experiments is a precise
knowledge of the nuclear matrix elements (NME) describing the half-life of
this process. 
We developed a new shell model approach for computing the $0\nu\beta\beta$ decay NME,
and we used it to analyze the $0\nu\beta\beta$ mode of $^{48}$Ca. 
The dependence of the NME on the short range correlations parameters, 
on the average energy of the intermediate states, on the finite-size cutoff parameters, 
and on the effective interaction used for the 
many-body calculations is discussed.
\end{abstract}

\maketitle

\section{Introduction}

The neutrinoless double beta ($0 \nu \beta \beta$) decay, which can only occur by violating the conservation of the total 
lepton number, if observed, will unravel physics beyond Standard Model (SM), and will represent 
a major milestone in the study of the fundamental properties of neutrinos \cite{HAX84}-\cite{rmp-08}.
Recent results from neutrino 
oscillation experiments have
convincingly demonstrated that neutrinos have mass and they can
mix \cite{kamland}-\cite{atmospheric}. The neutrinoless double beta ($0\nu\beta\beta$)
decay is the most sensitive process to determine the absolute
scale of the neutrino masses, and the only one that can
distinguish whether neutrino is a Dirac or a Majorana particle. A
key ingredient for extracting the absolute neutrino masses from
$0\nu\beta\beta$ decay experiments is a precise knowledge of the
nuclear matrix elements (NME) for this process. 
Since most of the $\beta\beta$ decay emitters are open shell nuclei,
many calculations of the NME have been performed within the pnQRPA approach and its 
extensions \cite{VOG86}-\cite{sim-97}. However, the pnQRPA calculations 
are very sensitive to the variation of the so called $g_{pp}$ parameter (the strength 
of the particle-particle interactions in the $1^+$ channel) \cite{VOG86}-\cite{SUH88}, and this 
drawback still persists in spite of various improvements brought by its 
extensions \cite{SMK90}-\cite{BKZ00}, including higher-order QRPA approaches \cite{RFS91}-\cite{sim-97}. 
The outcome of these attempts was that the calculations became more stable against $g_{pp}$ variation,
but 
at present there are still large differences between the values of the NME calculated with different 
QRPA-based methods, which do not yet provide a reliable determination of the 
two-neutrino double beta ($2\nu\beta\beta$) decay half-life.
Therefore, although the QRPA methods do not seem to be suited to predict the $2\nu\beta\beta$ 
decay half-lives, one
can use the measured $2\nu\beta\beta$ decay half-lives to calibrate the $g_{pp}$ parameters that are further used
to calculate the $0\nu\beta\beta$ decay NME \cite{rodin07}. 
Another method that recently provided NME for most $0\nu\beta\beta$ decay
cases of interest is the Interactive Boson Model \cite{iba-09}. Given the novelty of these
calculations, it remains to further validate their reliability by comparison with experimental data. 
  
On the other hand, the progress in computer power,  
numerical algorithms, and improved nucleon-nucleon effective interactions, made possible
large scale shell model calculations of the $2\nu\beta\beta$ and $0\nu\beta\beta$ decay
NME \cite{plb-ca48}-\cite{retamosa-95}.
The main advantage of the large scale shell model calculations is that they seem to be 
less dependent on the effective interaction used, as far as these interactions are consistent
with the general spectroscopy of the nuclei involved in the decay. Their main drawback is the 
limitation imposed by the exploding shell model dimensions on the size of the valence
space that can be used.
The most important success of the large scale shell model calculations 
was the correct prediction of the $2\nu\beta\beta$ decay half-life for $^{48}Ca$ \cite{plb-ca48,exp-ca48}. 
In addition, these calculations did not have to adjust any additional parameters, i.e.
given the effective interaction and the Gamow-Teller quenching factor extracted 
from the overall spectroscopy in the mass-region (including beta decay probabilities 
and charge-exchange form factors), one can reliably predict the $2\nu\beta\beta$ decay half-life of $^{48}Ca$.

Clearly, there is a need to further check and refine these calculations, and to provide
more details on the analysis of the NME that can be validated by experiments.
We have recently revisited \cite{HSB07} the two neutrino double beta decay of
$^{48}$Ca using two recently proposed effective interactions for this mass region,
GXPF1 and GXPF1A, and we explicitly analyzed the dependence of the double-Gamow-Teller
sum entering the NME on the excitation energy of the $1^+$ states in the 
intermediate nucleus $^{48}Sc$. This sum was recently investigated 
experimentally \cite{prl-0709}, and it was shown that indeed, the incoherent sum 
(using only absolute values of the Gamow-Teller matrix elements) would
provide an incorrect NME, validating our prediction.
We have also corrected by several orders of magnitude the probability of
transition of the g.s. of $^{48}$Ca to the first excited $2^{+}$ state of
$^{48}$Ti. Future experiments on double beta decay of $^{48}$Ca (CANDLES \cite{candles-06} and
CARVEL \cite{carvel-05}) may reach the required sensitivity of measuring such transitions, and
our results could be useful for planning these experiments.
\begin{figure}[tbe]
\centering
\includegraphics[width=0.46\textwidth]{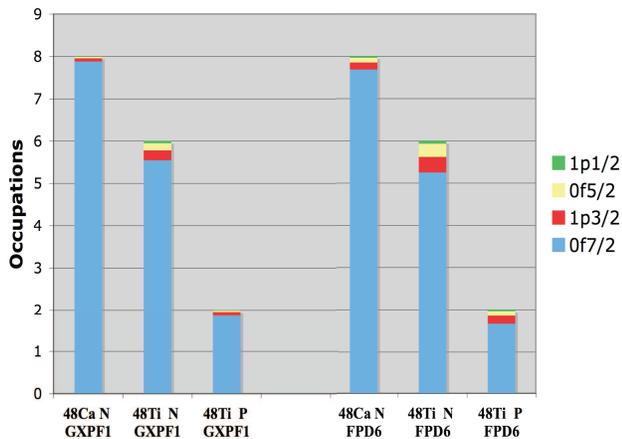}
\caption{(Color online) Comparison of the neutron and proton occupation probabilities between GXPF1 
interaction and FPD6 interaction.} 
\label{focc}
\end{figure}
\begin{table}[tbe]
	\caption{Neutron and proton occupation probabilities for the nuclei involved in the decay.} 
	\begin{center}
        \begin{tabular}{c|c|c|c|c|c}
            \hline 
        Nucleus (N/P) &    Interaction & $0f_{7/2}$ & $1p_{3/2}$ & $0f_{5/2}$ & $1p_{1/2}$\\
            \hline 
$^{48}$Ca N & GXPF1 & 7.883 & 0.073 & 0.033 & 0.011\\
$^{48}$Ti  N & GXPF1 & 5.545 & 0.237 & 0.167 &	0.051 \\
$^{48}$Ti  P & GXPF1 & 1.846 & 0.110 & 0.033 &	0.011\\
\hline
$^{48}$Ca N & GXPF1A	& 7.892 &	0.067 &	0.032 &	0.009\\
$^{48}$Ti  N & GXPF1A &	5.535 &	0.248 &	0.168 &	0.048\\
$^{48}$Ti  P & GXPF1A &	1.839 &	0.119 &	0.032 &	0.010\\
\hline
$^{48}$Ca N & KB3 &	7.800 &	0.0706 &	0.105 &	0.024\\
$^{48}$Ti  N & KB3 &	5.422 &	0.266	& 0.248 &	0.064\\
$^{48}$Ti  P & KB3 &	1.770	& 0.120 &	0.089 &	0.022\\
\hline
$^{48}$Ca N & KB3G &	7.795	& 0.070 &	0.112 &	0.024\\
$^{48}$Ti  N & KB3G &	5.416	& 0.260 &	0.263	& 0.061\\
$^{48}$Ti  P & KB3G &	1.763	& 0.120 &	0.097	& 0.021\\
\hline
$^{48}$Ca N &  FPD6 &	7.693	& 0.161 &	0.117	& 0.029\\
$^{48}$Ti  N & FPD6 &	5.253	& 0.369 &	0.310	& 0.068\\
$^{48}$Ti  P &  FPD6 &	1.673	& 0.196 &	0.101	& 0.031\\
            \hline
        \end{tabular}
	\end{center}
\label{tocc}
\end{table}

In the present paper we continue our investigations of the double beta decay of $^{48}$Ca,
analyzing the $0\nu\beta\beta$ decay NME. $^{48}$Ca has the largest $Q_{\beta \beta}$-value
of 4.271 MeV (next largest is that of $^{150}$Nd decay, 3.367 MeV), 
which could contribute to an increased decay probability. In addition, the high-energy gamma and
beta radiation emitted in this process could help eliminate most of the background noise.  
However, the small natural 
abundance of this isotope, 0.187\%, increases the difficulty of an experimental investigation, 
although new, improved separation techniques were recently proposed \cite{ogawa09}.
In addition, previous 
calculations \cite{retamosa-95,poves-07} suggest that its NME is a factor 4-5 smaller that those 
of other $\beta\beta$ emitters, such as $^{76}$Ge and $^{82}$Se \cite{prl100}.
Since these calculations were reported, it was shown that
the short range correlations might not have such a dramatic effect on the 
NME \cite{sim-09,eh-09} as previously thought,
and it was  also shown that higher order terms in the nucleon currents could be important \cite{sim-09}.
In the present paper we take into account all these new developments in the analysis of the 
NME for the $0\nu\beta\beta$ decay of $^{48}$Ca, and we study the NME dependence on the short range correlations 
parameters, on the finite size parameter, on the average energy of the intermediate states, 
and on the effective interaction used for the many-body calculations.

\section{The Neutrinoless Double Beta Decay Matrix Element} 

The $0 \nu \beta \beta$ decay $(Z,A) \rightarrow (Z+2,A)+2e^-$ requires the neutrino to be a 
Majorana fermion, i.e. it is identical to the antineutrino.  Considering only contributions from 
the light Majorana neutrinos \cite{rmp-08}, the $0\nu\beta\beta$ decay half-live is given by

\begin{equation}
\left( T^{0\nu}_{1/2} \right)^{-1}=G_1^{0\nu}\mid M^{0\nu}\mid^2 
\left( \frac{ \left< m_{\beta \beta} \right> }{m_e} \right)^2 \ ,
\end{equation}

\noindent
Here, $G_1^{0\nu}$ is the phase space factor, which depends on the $0\nu\beta\beta$ decay energy,
$Q_{\beta\beta}$  and the nuclear radius. 
The effective neutrino mass, $\left< m_{\beta \beta} \right>$, is related to the neutrino mass eigenstates, 
$m_k$, via the lepton mixing matrix, $U_{e k}$,

\begin{equation}
\left< m_{\beta \beta} \right> = \mid \sum_{k} m_{k} U^{2}_{e k} \mid \ .
\end{equation}

\noindent
The NME, $M^{0\nu}$, is given by
\begin{equation}
 M^{0 \nu}=M^{0 \nu}_{GT}-\left( \frac{g_V}{g_A} \right)^2  M^{0 \nu}_F- M^{0 \nu}_T\ ,
\end{equation}

\noindent
where $M^{0 \nu}_{GT}$, $M^{0 \nu}_F$ and $M^{0 \nu}_T$ are the Gamow-Teller (GT), 
Fermi (F) and tensor (T) matrix elements, 
respectively. These matrix elements are defined as follows:

\begin{equation}
M_\alpha^{0\nu} = \sum_{m,n} \left< 0^+_f \mid \tau_{-m} \tau_{-n} O^\alpha_{mn} \mid 0^+_i \right> \ ,
\end{equation}

\noindent
where $O^\alpha_{mn}$ are $0\nu\beta\beta$ transition operators, $\alpha=(GT,\ F,\ T)$, $\mid 0^+_i >$
is the g.s. of the parent nucleus (in our case $^{48}$Ca), and  $\mid 0^+_i >$ is the g.s. of the
grand daughter nucleus (in our case $^{48}$Ti).
\\ \\
Due to the two-body nature of the transition operator, the matrix element can be  
reduced to a sum of products 
of two-body transition densities TBTD and antisymmetrized two-body matrix elements, 

\begin{eqnarray}
\nonumber M_\alpha^{0\nu} & = & \sum_{j_p j_{p^\prime} j_n j_{n^\prime} J_\pi} TBTD 
\left( j_p j_{p^\prime} , j_n j_{n^\prime} ; J_\pi \right) \\ 
& & \left< j_p j_{p^\prime}; J^\pi T \mid \tau_{-1} \tau_{-2}O^\alpha_{12} 
\mid j_n j_{n^\prime} ; J^\pi T\right>_a ,
\label{mbme}
\end{eqnarray}
where $O^\alpha_{12}$ are given by
\begin{eqnarray}
\nonumber O_{12}^{GT} = \vec{\sigma}_1 \cdot \vec{\sigma}_2 H_{GT}(r) \ , \\
\nonumber O_{12}^{F} = H_{F}(r)\ ,  \\ 
O_{12}^{T} = \left[3\left(\vec{\sigma}_1 \cdot \hat{r} \right) 
\left(\vec{\sigma}_1 \cdot \hat{r} \right) - 
\vec{\sigma}_1 \cdot \vec{\sigma}_2 \right] H_{T}(r)\ .
\end{eqnarray}

The matrix elements of $O^\alpha_{12}$ for the $jj$-coupling scheme consistent with the
conventions used by modern shell model effective interactions are described
in the Appendix.

\begin{table}[tbe]
	\caption{Parameters for the SRC parametrization of Eq. (\ref{fsrc}).} 
	\begin{center}
        \begin{tabular}{c|c|c|c}
            \hline 
      SRC       & $a$ & $b$ & $c$ \\
            \hline 
Miller-Spencer & 1.10 & 0.68 & 1.00 \\
CD-Bonn & 1.52 & 1.88 & 0.46 \\ 
AV18  & 1.59 & 1.45 & 0.92\\
            \hline
        \end{tabular}
	\end{center}
\label{tsrc}
\end{table}

\begin{figure}[tbe]
\centering
\includegraphics[width=0.46\textwidth]{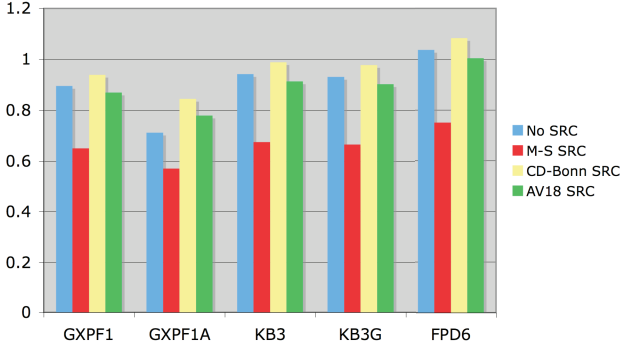}
\caption{(Color online) The dependence of the NME on the effective interaction used and the short 
range correlation (SRC) model. M-S stands for Miller-Spencer.} 
\label{fnme}
\end{figure}

To calculate the two-body matrix elements in Eq. (5) one needs the neutrino potentials 
entering into the radial matrix element $\left< nl \mid H_\alpha \mid nl^\prime \right>$ in 
Eq. (\ref{tbme}) below.
Following Ref. \cite{sim-09} and using closure approximation one gets,

\begin{eqnarray}
H_\alpha(r) & = & \frac{2R}{\pi}\int^\infty_0 f_{\alpha}(qr)
\frac{h_\alpha(q^2)}{q+\left< E \right> }G_\alpha(q^2) q dq \ ,
\end{eqnarray}
\noindent
where $f_{F,GT}(qr)=j_0(qr)$ and $f_T(qr)=j_2(qr)$ are spherical Bessel functions, 
$\left< E \right>$ is the average energy of the virtual intermediate states used in the closure approximation,
and the form factors $h_\alpha(q^2)$ that include the higher order terms in the nucleon currents are

\begin{eqnarray}
\nonumber 
h_F(q^2) & = & g_V^2(q^2) \\
\nonumber
h_{GT}(q^2) & = & \frac{g_A^2(q^2)}{g_A^2}
\left[1-\frac{2}{3}\frac{q^2}{q^2+m_\pi^2}+\frac{1}{3}\left(\frac{q^2}{q^2+m_\pi^2}\right)^2\right] \\
\nonumber
   & & +\frac{2}{3}\frac{g_M^2(q^2)}{g_A^2}\frac{q^2}{4m_p^2}\ ,\\
\nonumber
h_T(q^2) & = &  \frac{g_A^2(q^2)}{g_A^2}
\left[\frac{2}{3}\frac{q^2}{q^2+m_\pi^2}-\frac{1}{3}\left(\frac{q^2}{q^2+m_\pi^2}\right)^2\right] \\
   & & +\frac{1}{3}\frac{g_M^2(q^2)}{g_A^2}\frac{q^2}{4m_p^2}\ .
\label{npots}
\end{eqnarray}

The $g_{V,A,M}$ form factors in Eq.(\ref{npots}) can include nucleon finite size effects, which in the dipole 
approximation are given by

\begin{eqnarray}
\nonumber
g_V(q^2)=\frac{g_V}{\left(1+q^2/\Lambda_V^2\right)^2}, \\
\nonumber
g_M(q^2)=(\mu_p-\mu_n) g_V(q^2), \\
g_A(q^2)=\frac{g_A}{\left(1+q^2/\Lambda_A^2\right)^2}.
\end{eqnarray}

\noindent
Here $g_V=1$, $g_A=1.25$, $(\mu_p-\mu_n)=3.7$, $\Lambda_V=850\ MeV$, and $\Lambda_A=1086\ MeV$.

\begin{table}[tbe]
	\caption{Different contributions to the NME for the GXPF1A interaction with $\left< E \right> =7.72 \ MeV$} 
	\begin{center}
        \begin{tabular}{c|c|c|c|c}
            \hline 
      SRC       & $M^{0 \nu}_{GT}$ & $M^{0 \nu}_F$ & $M^{0 \nu}_T$ & $M^{0 \nu}$\\
            \hline 
None & 0.556 & -0.219 & -0.015 & 0.711 \\
Miller-Spencer & 0.465 & -0.141 & -0.014 & 0.570 \\
CD-Bonn & 0.688 & -0.222 & -0.014 & 0.845 \\ 
AV18  & 0.634 & -0.204 & -0.014 & 0.779 \\
            \hline
        \end{tabular}
	\end{center}
\label{tnme}
\end{table}

The short range correlations are included via the correlation function $f(r)$ that modifies 
the relative wavefunctions at short distances,

\begin{equation}
\psi_{nl}(r) \rightarrow \left[ 1+f(r) \right] \psi_{nl}(r) \ ,
\end{equation}

\noindent
where $f(r)$ can be parametrized as \cite{sim-09},
\begin{eqnarray}
f(r) = -c e^{-ar^2} \left( 1-br^2 \right) \ .
\label{fsrc}
\end{eqnarray}

\begin{figure}[tbe]
\centering
\includegraphics[width=0.46\textwidth]{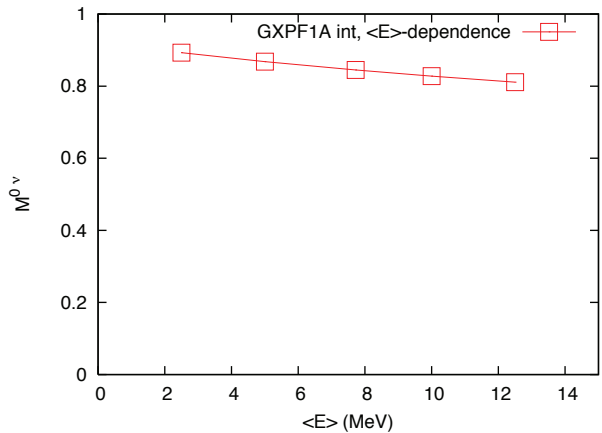}
\caption{(Color online) The dependence of the NME on the average energy 
of the intermediate states, $\left< E \right>$, for the GXPF1A interaction.} 
\label{fmve}
\end{figure}

The radial matrix elements of $H_\alpha$ between relative harmonic oscillator wavefunctions $\psi_{nl}(r)$ and 
$\psi_{n^\prime l^\prime}(r)$, 
$\left\langle nl \mid {H}_\alpha (r) \mid n^\prime l^\prime \right\rangle$,  become

\begin{eqnarray}
\int^\infty_0 r^2 dr \psi_{nl}(r) H_\alpha(r) \psi_{n^\prime l^\prime}(r)\left[ 1+f(r) \right]^2 
\end{eqnarray}


Although the neutrino potentials are quite close to a Coulomb potential, the integrands in
Eq. (7) are strongly oscillating,  and the integrals require special numerical treatment. 

Having calculated the two body matrix elements, we developed a new shell model approach for computing the 
many-body matrix elements for $0\nu\beta\beta$ transition, Eq. (5). This approach is briefly described in 
the Appendix.

\section{Results}

In this study of the $0\nu\beta\beta$ decay NME we used five different effective interactions 
available for the shell model description of the  $pf$-shell nuclei: GXPF1\cite{gxpf1},
GXPF1A \cite{gxpf1a}, KB3 \cite{kb3}, KB3G \cite{kb3g}, and FPD6 \cite{fpd6}.
These effective interaction were constructed starting from a G-matrix \cite{rg} in the $pf$ shell,
which was further adjusted
 to describe some specific (but different) sets of experimental energy levels
of some $pf$-shell nuclei. Although their elements are quite different, their predictions of the
spectroscopic observables around A=48 are not very far apart. Recent experimental 
investigations \cite{scf-prl,scf-prc}
of the nucleon occupation probabilities in $^{76}$Ge and $^{76}$Se and the subsequent 
theoretical analysis \cite{sim-09,poves-09} highlighted the relevance of these observables 
to obtain an accurate
description of the nuclear structure of the nuclei involved in the double beta decay.
Fig. \ref{focc} compares the neutron and proton occupation probabilities in $^{48}$Ca
and $^{48}$Ti for two different effective interactions, GXPF1 and FPD6. One can see very
small differences between the results of the two interactions.  One
can come to the same conclusion when comparing the similar occupations probabilities for all 
five interactions given in Table \ref{tocc}.

\begin{figure}[tbe]
\centering
\includegraphics[width=0.46\textwidth]{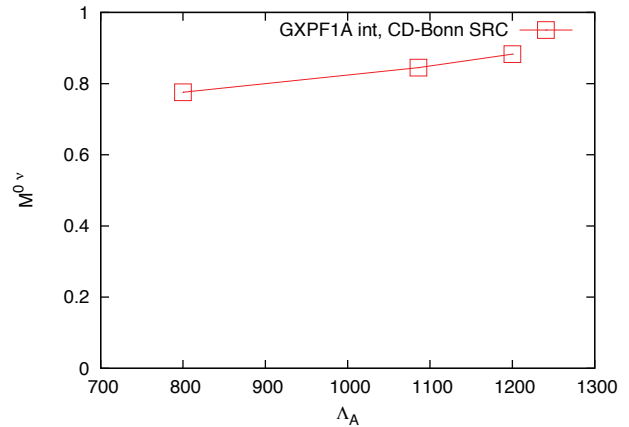}
\caption{(Color online) The dependence of the NME on the axial cutoff parameter
$\Lambda_A$ for the GXPF1A interaction.} 
\label{fmvl}
\end{figure}

In the present calculations we considered both  short range 
correlation (SRC) effects and finite size (FS) effects. 
Although the radial dependence of the neutrino potential is very close to that of a Coulomb potential,
many previous calculations \cite{CPZ90,retamosa-95,poves-07,prl100} took into account the SRC missing in the two-body 
product wave functions, via the Jastrow-like parametrization described in Eqs. (10)-(12).
Until recently, the parameters $a$, $b$, $c$ used were those proposed by Miller and Spencer \cite{src-ms},
which have the effect of decreasing the NME by about 30\%. Recently, \cite{sim-09}
the SRC effects were revisited, using modern nucleon-nucleon interactions, such as 
CD-Bonn and AV18, and it was found that the decrease of the relative wave functions at
short distances is compensated by a relative increase at larger distances, and the overall
NME do not change very much when compared with the NME without SRC effects.
Ref. \cite{sim-09} proposed a parametrization of these results in terms of
similar Jastrow-like correlation functions as in Eq. (10)-(11), the corresponding
parameters being listed in Table \ref{tsrc}. In addition, Ref. \cite{eh-09} introduced an
effective $0\nu\beta\beta$ operator that takes into account the SRC effects and the contribution 
of the missing shells from the valence space using the general theory of 
effective interactions \cite{rg}, and found that the NME for the $0\nu\beta\beta$ decay of
$^{82}$Se did not change significantly when compared with the result of the "bare"
operator.

Fig. \ref{fnme} shows our NME for all five effective interactions, for all three SRC
set of parameters given in Table \ref{tsrc}, as well as for no SRC. One can see that
the semi-quantitative analysis above is reflected in the dependence of NME on the
choice of SRC. The results do not show significant dependence on the effective interaction
used, although one can see a 20\% spread of NME for the same choice of SRC.
All NME reported here contain the higher order terms described in Eqs. (7)-(9).
A comparison with the NME calculated without the higher order terms in the potential will be reported
elsewhere. To be consistent \cite{cowell} with the calculation of the phase factor, 
$G^{0\nu}_1$, we used $R=1.2 A^{1/3}\ fm$ in Eq. (7). Our choice for $\hbar \omega$ 
parameter entering the harmonic oscillator wave functions was
$45 A^{-1/3} - 25 A^{-2/3}$, which was shown to provide better shell model
description of observables than the simple $41 A^{-1/3}$ anstaz. Table \ref{tnme}
shows the GT, F, and T contributions to the overall NME for all SRC choices, 
when the GXPF1A interaction was used. One can see that all theses contributions add
coherently in Eq. (3), and that the tensor contribution is negligible in all cases.

Fig. \ref{fmve} shows the dependence of the NME of the average energy of the intermediate
states. Varying $\left<E\right>$ from 2.5 MeV to 12.5 MeV one gets less than 5\% variation
in the NME. This result suggests that closure approximation is quite good, although
a direct study might be necessary to find out the exact magnitude of the error. All other
NME results reported here used $\left<E\right>= 7.72$ MeV.
Fig. \ref{fmvl} shows the dependence of NME of the axial cutoff parameter $\Lambda_A$, while
$\Lambda_V$ is kept fixed at 850 MeV. The variation is within 5\%, indicating a 
weak dependence of the FS cutoff parameters.
The FS effects were implemented via the cutoff parameters $\Lambda_V$
and $\Lambda_A$ in the form factors given in Eq. (9). In most of our results, except those in
Fig. \ref{fmvl}, we use the same FS cutoff parameters as in Refs. \cite{rmp-08,sim-09}, i.e. 
$\Lambda_V=850$ MeV and $\Lambda_A=1086$ MeV. In Fig. \ref{fmvl} we present the dependence
of the NME on $\Lambda_A$, while $\Lambda_V$ is kept at its nominal value. As with the $\left<E\right>$
dependence, the results varies by less than 5\%.

\section{Conclusions and Outlook}

In conclusion, we presented a new shell model analysis of the nuclear matrix elements for the neutrinoless 
double beta decay  of $^{48}$Ca. We included in the calculations the recently proposed higher order terms
of the nucleon currents, three old and recent parametrization of the SRC effect,
FS effects, intermediate states energy effects, and we treated careful few other 
parameters entering the into the calculations.
We found very small variation of the NME with the average energy of the intermediate states, 
and FS cutoff parameters, and moderate variation vs the effective interaction and SRC
parametrization. Our overall average NME using all values presented in Fig. \ref{fnme}
is 0.86, although the elimination of the Miller-Spencer SRC parametrization would significantly increase
this value. We estimate the error due to the effects studied here to be about 18\%.
Using the present value of the NME and the recommended \cite{cowell} phase-space factor 
$G^{0\nu}_1 = 6.5\times 10^{-14}\ yr^{-1}$ one can conclude that a future measurement of
the $0\nu\beta\beta$ decay half-life of $10^{26}\ yr$, which seems to be the limit
imposed by the present energy resolution of the CANDLES detector \cite{ogawa09} (see also Fig. 21 of 
Ref. \cite{avignone-05}), could detect a neutrino mass $\left< m_{\beta\beta} \right>$
of about $230\pm45$ meV. New improvements in the detector technology could further reduce this limit.

We believe that our analysis covered the most important effects relevant to the accuracy of
the NME for the double beta decay of $^{48}$Ca. The successful prediction of the $2\nu\beta\beta$ decay
half-live of $^{48}$Ca using the shell model approach in the same model space suggests that 
the $0\nu\beta\beta$ decay half-life could be reliably predicted. 
Our analysis indicates that closure approximation
is accurate at the level of 5\% error. However, a direct comparison with the sum on intermediate
states, calculated within the shell model, would be more reassuring. 

As in all other $0\nu\beta\beta$ decay calculations reported so far, no quenching factors have been used.
This is probably the least investigated potential effect on the $0\nu\beta\beta$ decay NME.   
It is possible that the GT-like operators be quenched even if they have finite range.
No definitive conclusion exists about these effects, although
Ref. \cite{eh-09} seems to indicate that they might not play a major role. 
One could try getting some insight into this
problem by investigating the fictitious $0\nu\beta\beta$ decay of a light nucleus, such as $^{16}$Be,
and increasing the valence space one can study the changes of different contributions
to the NME.

\section{Appendix}

\noindent
The matrix element of $O^\alpha_{12}$ for non-antisymmetrized $jj$-coupling 
two-particle states can be decomposed into products of reduced matrix elements of the spin, 
relative and center of mass operators,

\begin{widetext}
\begin{eqnarray}
\nonumber & & \left\langle n_1 l_1 j_1, n_2 l_2 j_2 ; J^\pi T=1\mid O^\alpha_{12} 
\mid n^\prime_1 l^\prime_1 j^\prime_1, n^\prime_2 l^\prime_2 j^\prime_2 ; J^\pi T=1\right\rangle =  \\
\nonumber & & \sum_{S \lambda} \left\langle  l_1\frac{1}{2}(j_1), 
l_2\frac{1}{2}(j_2) \mid \frac{1}{2}\frac{1}{2} (S) , l_1 l_2 (\lambda) 
\right\rangle _J \left\langle l^\prime_1\frac{1}{2}(j^\prime_1), 
l^\prime_2\frac{1}{2}(j^\prime_2) \mid \frac{1}{2}\frac{1}{2} (S) , 
l^\prime_1 l^\prime_2 (\lambda) \right\rangle _J 
 \frac{1}{\sqrt{ \left( 2S+1 \right) }} \left\langle \frac{1}{2}\frac{1}{2};
S \| S^{(0)}_\alpha \| \frac{1}{2}\frac{1}{2};S \right\rangle   \\
& & \times  \sum_{nn^{\prime}lNL} \left< nl,NL \mid n_1 l_1, n_2 l_2 \right> _\lambda 
\left< n^{\prime}l,NL \mid n^{\prime}_1 l^{\prime}_1, n^{\prime}_2 l^{\prime}_2 
\right> _\lambda \left< nl \mid H_{\alpha} (r) \mid n^{\prime}l \right> .
\label{tbme}
\end{eqnarray}

\noindent
where

\begin{equation}
\left\langle  l_1\frac{1}{2}(j_1), l_2\frac{1}{2}(j_2) \mid \frac{1}{2}\frac{1}{2} (S) , 
l_1 l_2 (\lambda) \right\rangle _J = \sqrt{(2j_1+1)(2j_2+1)(2S+1)(2 \lambda +1)} \left( \begin{matrix}
 l_1 & \frac{1}{2} &  j_1 \\
 l_2 & \frac{1}{2} &  j_2 \\
 \lambda & S &  J
\end{matrix}
\right) \ ,
\end{equation}
\end{widetext}

\noindent
and the last factor is a 9j symbol. As mentioned in the text, the order of the spin-orbit
coupling must be consistent with the convention considered into the derivation of the
effective interaction used for the many body calculations ($\vec{l}+\vec{s}$ in most cases). 
The reduced spin matrix elements are

\begin{eqnarray}
\nonumber
\left\langle  \frac{1}{2} \frac{1}{2} S \| \vec{\sigma}_1 \cdot \vec{\sigma}_2 \| \frac{1}{2}\frac{1}{2} 
S \right\rangle & = & \sqrt{2S+1}\left[2S(S+1)-3\right] \ , \\
\left\langle  \frac{1}{2} \frac{1}{2} S \| 1 \| \frac{1}{2}\frac{1}{2} S \right\rangle & = & 
\sqrt{2S+1} \ .
\end{eqnarray}

The expressions for the matrix elements of the tensor operator will be given elsewhere. Our calculations 
show that the tensor term gives a small contribution to the NME in most cases. 
The higher order terms in the nucleon currents,
however, decrease the overall NME by about 20-25\%.
The antisymmetrized form of the two-body matrix elements can be obtained using

\begin{eqnarray}
\nonumber \left< j_p j_{p^\prime}; J^\pi T \mid t_{-1} t_{-2}O^\alpha_{12} 
\mid j_n j_{n^\prime} ; J^\pi T\right>_a= \\
\nonumber
\frac{1}{\sqrt{(1+\delta_{j_p j'_{p^\prime}})(1+\delta_{j_n j_{n^\prime}})}}\times \\
\nonumber [\left< j_p j_{p^\prime}; J^\pi T \mid t_{-1} t_{-2}O^\alpha_{12} 
\mid j_n j_{n^\prime} ; J^\pi T\right> -\\
(-1)^{j_n+j_{n^\prime}+J}
\left< j_p j_{p^\prime}; J^\pi T \mid t_{-1} t_{-2}O^\alpha_{12} 
\mid j_{n^\prime} j_n ; J^\pi T\right> ]
\end{eqnarray}

Having the two-body matrix elements ready, one can calculate the NME in Eq. (\ref{mbme}) 
 if two body transition densities 
$TBTD\left( j_p j_{p^\prime} , j_n j_{n^\prime} ; J_\pi \right)$
are known. Most of the shell model codes do not provide two body transition densities.
One alternative approach is to take advantage of the isospin symmetry that most
of the effective interactions have, which creates wave functions with good isospin. 
The approach described
below works also when the proton and neutron are in different shells.
If the above conditions are satisfied, one can transform the two body matrix elements of
a $\Delta T=2$ operator using the Wigner-Eckart theorem, from $\Delta T_z=-2$ to 
 $\Delta T_z=0$, which can be further used (see below) to describe transitions between states of 
the same nucleus. Denoting

\begin{eqnarray}
\left< O^{\Delta T=2}_{\Delta T_z =-2} \right> = \\
\left< T=1\ T_z=-1 \mid O^{\Delta T=2}_{\Delta T_z =-2} \mid T=1\ T_z=1 \right>,
\end{eqnarray}

\noindent
one gets for the $\Delta T_z=0$ the two body matrix elements

\begin{eqnarray}
\nonumber
< pp \mid O^{\Delta T=2}_{\Delta T_z =0} \mid pp > = \\
\nonumber
< nn \mid O^{\Delta T=2}_{\Delta T_z =0} \mid nn > = \\
< O^{\Delta T=2}_{\Delta T_z =-2} > \times C^{1\ 2\ 1}_{1\ 0\ 1}/C^{1\ \ \ 2\ \ 1}_{1\ -2\ -1} ,
\label{menn}
\end{eqnarray}

\noindent
and

\begin{eqnarray}
\nonumber
< pn\ T=1 \mid O^{\Delta T=2}_{\Delta T_z =0} \mid pn\ T=1 > = \\
< O^{\Delta T=2}_{\Delta T_z =-2} > \times C^{1\ 2\ 1}_{0\ 0\ 0}/C^{1\ \ 2\ \ 1}_{1\ -2\ -1} ,
\label{mepn}
\end{eqnarray}

\noindent
where $C^{1\ \ \ 2\ \ 1}_{T_{zi}\ \Delta T_z\ T_{sf}}$ are isospin Clebsch-Gordan coefficients.
The transformed matrix elements in Eqs. (\ref{menn}) and (\ref{mepn}) preserve spherical 
symmetry and they can be used as a piece of a Hamiltonian, $H^\alpha_{\beta \beta}$ ,
that violates isospin symmetry. 

One can then lower by 2 units the isospin projection of the g.s. of the parent nucleus that 
has the higher isospin $T_{>}$, $^{48}$Ca in our case, thus becoming an isobar analog state
of the grand daughter nucleus that has isospin $T_{<} = T_{>}-2$, $^{48}$Ti in our case. Denoting
by $\mid 0^{+}_{i<}\ T_> >$ the transformed state, one can now calculate the many body matrix 
elements of the transformed $0\nu\beta\beta$ operator,

\begin{equation}
M_\alpha^{0\nu}(T_z=T_<)=\left< 0^+_f\ T_< \mid H^\alpha_{\beta \beta} \mid 0^+_{i<}\ T_> \right> .
\label{metrd}
\end{equation}

\noindent
Choosing $\mid 0^{+}_{i<}\ T_> >$ as a starting Lanczos vector, and performing one Lanczos
iteration with $H^\alpha_{\beta \beta}$ one gets

\begin{equation}
H^\alpha_{\beta \beta} \mid 0^{+}_{i<}\ T_> > = a_1 \mid 0^{+}_{i<}\ T_> > + b_1 \mid L_1 > ,
\end{equation}

\noindent
where $\mid L_1 >$ is the new Lanczos vector. Then, one can calculate the matrix elements
in Eq. (\ref{metrd}) 

\begin{equation}
M_\alpha^{0\nu}(T_z=T_<)=b_1 \left < 0^+_f\ T_< \mid L_1 \right> .
\label{metr}
\end{equation}

The transition matrix elements in Eq. (\ref{mbme}) can then be recovered using again the 
Wigner-Eckart theorem,

\begin{equation}
M_\alpha^{0\nu}=M_\alpha^{0\nu}(T_z=T_<)\times C^{T_>\ \ 2\  T_<}_{T_>\ -2\ T_<}/
C^{T_>\ 2\ T_<}_{T_<\ 0\ T_<} .
\end{equation}

Although it looks complicated, this procedure is rather easy to implement. The transformation 
of the g.s. of the parent to an analog state of the grand daughter can be performed very quickly,
and one Lanczos iteration represents a small load as compared with the calculation of the
g.s. of the grand daughter. The additional calculations described in Eqs. (22)-(24) require 
smaller resources than those necessary to calculate the TBTDs.


\begin{acknowledgments}
M.H. acknowledges support from NSF Grant No. PHY-0758099.
\end{acknowledgments}

\end{document}